# Enhancement of flux pinning and high critical current density in graphite doped MgB$_2$ superconductor

Chandra Shekhar[a)], Rajiv Giri, R. S. Tiwari and O. N. Srivastava[a)]
*Department of Physics Banaras Hindu University, Varanasi-221005, India*
S. K. Malik
*Tata Institute of Fundamental Rresearch, Mumbai-400005, India*

**ABSTRACT**

We report the synthesis and characterization of graphite (C) doped MgB$_{2-x}$C$_x$ (x = 0.0, 0.1, 0.2 and 0.3) samples. The crystal structure and microstructural characterization have been investigated by x-ray diffractometer and transmission electron microscopic (TEM) analysis. The superconducting properties especially $J_c$ and $H_{c2}$ have been measured by employing physical property measurement system. We found that the graphite doping affects the lattice parameters as well as the microstructure of MgB$_2$ superconductor. In case of optimally doped (x = 0.1) sample, the critical current density at 5K corresponds to 1.1 x 10$^6$ and 5.3 x 10$^4$ A/cm$^2$ for 3T and 5T fields respectively. The upper critical field has been enhanced nearly two times after doping. The flux pinning behavior has been investigated by flux pinning force density curve and it reveals that the flux pinning behaviour has improved significantly by doping. TEM micrographs show the graphite nanoparticles of size ~5-10 nm which are invariably present in MgB$_2$ grains. These nanoparticles act as flux pinning centre and are responsible for enhancement of superconducting properties of MgB$_2$.

--------------------------------------------------------

[a)]e-mail: hepons@yahoo.com ( O N Srivastava), chand_bhu@yahoo.com (Chandra Shekhar)
Phone & Fax: +91 542 2369889

## INTRODUCTION

After the discovery of superconductivity in MgB$_2$ [1], considerable effort has been made to improve the critical current density ($J_c$), upper critical field ($H_{c2}$) and irreversibility field ($H_{irr}$) of this superconductor. The absence of weak links at grain boundaries in MgB$_2$ [2,3] makes it easier to improve superconducting abilities by introducing additional pinning centres in MgB$_2$ superconductor. Doping of elements or compounds has been found to be effective for improving the superconducting properties of bulk, tapes, wires and films especially under magnetic fields. The enhanced properties of above mentioned form of MgB$_2$ superconductor are well above those of standard high field materials e.g. Nb-based superconductor [4]. This raises the possibility of using MgB$_2$ as replacement of Nb-based superconductors. Large number of dopants e.g. carbon [5-11] as well as carbon containing compounds, SiC [12-19], B$_4$C [20-22] carbohydrate [23] and aromatic hydrocarbon [24] have been reported to be effective for enhancement of superconducting properties such as $H_{irr}$, $H_{c2}$ and $J_c$ under higher magnetic field. In these cases, doping of carbon and carbon containing compound resulted in the substitution of carbon at boron site and introduction of nonsuperconducting particles, which provide effective flux pinning centres and resulted in significant enhancement of $H_{c2}$ and $J_c$ [25-29]. Reduction of coherence length due to enhanced impurity scattering is considered to contribute to the enhancement of $H_{c2}$ [30-33]. Furthermore, flux pinning strength was found to be enhanced by carbon substitution [6,19,20,23,24]. These positive effects of carbon substitution indicate that doping of carbon has a great potential to enhance superconducting properties of MgB$_2$ superconductor. Although most of earlier investigations are related to the effect of carbon substitution on the superconducting properties of MgB$_2$, relatively sparse studies have been carried out on the microstructures of doped MgB$_2$ [10,11]. However, the arrangement of carbon atoms in graphite (C) sheet is somewhat similar to the arrangement of B atoms in MgB$_2$, studies on graphite doping is of special significance. The earlier reports present only comparison of SiC, C$_{60}$, CNT and graphite doping on J$_c$ in MgB$_2$ [34,35]. In order to enhance the value of $J_c$ and $H_{c2}$, however, investigation pertaining to optimization of size of carbon nanoparticles and their homogeneous distribution in the superconductor would be required. In the present work the effect of graphite doping on superconducting properties like, $T_c$, $J_c$, $H_{c2}$ and $H_{irr}$ of MgB$_2$ superconductor have been carried out. We have studied the effect of graphite doping on the crystal lattice, superconducting properties and their correlation with microstructures of MgB$_2$ superconductor which have been prepared by encapsulation method developed by us [36,37]. In the present work, we have evaluated $T_c$, $J_c$, $H_{c2}$, $H_{irr}$ and bulk flux pinning force density ($F_p$) from magnetization measurement of undoped and doped MgB$_2$ superconductors. The structural and microstructural properties have been carried out employing powder x-ray diffraction (XRD) and transmission electron microscopy (TEM) technique in diffraction and imaging modes.

## EXPERIMENTAL DETAILS

The synthesis of graphite doped MgB$_2$ samples with stoichiometric ratio MgB$_{2-x}$C$_x$ (x = 0.0, 0.1, 0.2 and 0.3) have been carried out by solid state reaction method at ambient pressure by employing a special encapsulation technique [36,37] developed in our laboratory. Magnesium Mg (purity - 99.9%, size - 30-40µm), amorphous boron B (purity - 99%, size - 5-6µm) and graphite C ( purity - 99.9%, size 40-50 nm) were fully mixed and cold pressed into small rectangular pellets (10x5x1) mm$^3$. The pellets were encapsulated with Mg metal cover to take care of Mg loss and avoid the formation of MgO during sintering process. The pellet configuration was rapped in a Ta foil and sintered in flowing high purity Ar gas in a programmable tube type furnace at 900$^0$C for 2 hours. The pellets were cooled to room temperature at the rate of 5$^0$C /min. The encapsulating Mg cover was removed and pellets were retrieved for further study.

All the samples were subjected to crystal structure characterization by powder x-ray diffraction technique (XRD, PANalytical X' Pert Pro, CuK$_\alpha$ radiation with $\lambda$ = 1.5406 Å) and microstructural characterization by transmission electron microscopy (TEM, Philips-CM-12). Magnetization measurements have been done by physical property measurement system (PPMS, Quantum Design, TIFR Mumbai, India ) on pellet of diameter and length 1.2 mm 4 mm respectively of as synthesized MgB$_2$ samples over a temperature range of 5-50K. The J$_c$ of pellet samples was calculated by using Bean's formula based on critical state model [38].

$$J_c = \frac{30\Delta M}{\langle d \rangle}$$

where '$\Delta M$' (emu/cm$^2$) is the height of hysterisis loop and $\langle d \rangle$ (cm) is the diameter of pellet which is used in magnetization measurement.

**RESULTS & DISCUSSION**

Fig (1) shows XRD patterns of MgB$_{2-x}$C$_x$ samples for x = 0.0, 0.1, 0.2 and 0.3. Excepting peaks at 26.2$^0$ and 52.4$^0$, all peaks are identified by MgB$_2$ compound with space group P6/mmm. These additional two peaks are identified as (002) and (004) reflections of graphite. The intensity of these peaks increases with doping as shown in Fig. 1(b, c & d) and no other impurity phase such as MgB$_2$C$_2$ has been found [21]. This XRD pattern reveals that the undoped and doped MgB$_2$ samples are polycrystalline in nature. XRD analysis using a computerized programme based on least square fitting method gives lattice parameters $a$ = 3.078 Å and $c$ = 3.522Å for pure MgB$_2$ sample. It is very close to the standard values [39]. We have noticed the shift in the positions of (100) and (002) peaks corresponding to MgB$_2$ phase with increasing concentration of graphite. The position of the (100) peak is shifted to higher angles with increasing level of doping, indicating a decrease in the '$a$' lattice parameter as shown in Fig. 2(a). However, the position of the (002) peak has only slightly changed with increasing doping [9,10,30] as shown in Fig. 2(b). As can be seen in Fig 2(c), the in-plane lattice parameter '$a$' decreases from 3.078 to 3.051Å and lattice parameter '$c$' from 3.522 to 3.523Å for doping x = 0.2. This can be understood because the average size of C atom ($r_C$ = 0.772Å) is smaller than that of B atom ($r_B$ = 0.822Å). The change in lattice parameters indicates that C atom is substituted in the boron honeycomb layer in the MgB$_2$ crystal. Furthermore, doping was also confirmed by analysis of full width at half maximum (FWHM) values for peaks (100), (101), (002) and (110) as shown in Fig 2(d). The broadening in diffraction line has also increased with doping. Broadening in peaks in present investigation is likely to arise from lattice strain mainly caused by doping on B sites. It may be noted that decrease in grain size could also result in peak broadening. However, by SEM investigation, the grain size seemed to remain the same for all samples.

The magnetic susceptibility ($\chi$) of MgB$_{2-x}$C$_x$ (x = 0.0, 0.1, 0.2 and 0.3) samples are shown in Fig. (3) as a function of temperature. Based on this, the transition temperature of pure and doped MgB$_2$ with different concentration can be taken to lie between 34-40K. The $T_c$ decreases with increasing doping concentration. Most importantly, the pure MgB$_2$ has $T_c$ at 40K while the doped material has $T_c$ ranging from 36-34 K. Thus doped samples show somewhat lower but still sharp $T_c$. The $T_c$ drops only 4K [21] for x = 0.1 suggesting that only a small amount of C atom has substituted at B site in our samples.

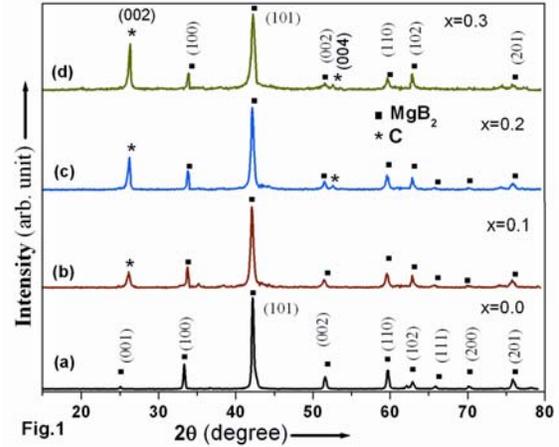

*FIG.1. (Color) Representative powder XRD patterns of MgB$_{2-x}$C$_x$ (a) x = 0.0, (b) x = 0.1, (c) x = 0.2 and (d) x = 0.3.*

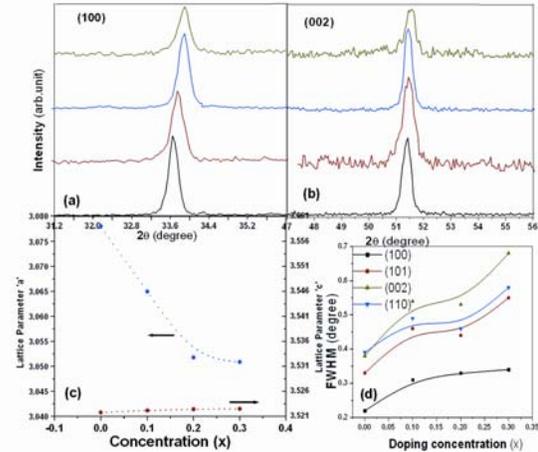

*FIG.2. (Color) (a) & (b) show the magnified view of the XRD patterns corresponding to (100) and (002) reflections respectively (c) shows change in lattice parameters with doping concentration (d) shows the variation in FWHM of (100), (101), (002) and (110) reflections with doping concentration.*

the magnetization measurements as function of applied magnetic field have been carried out at 5, 10, 20 and 30 K for each sample. The dependence of $J_c$ on the applied magnetic field is shown in Fig. (4). It is clear from figure that the $J_c$ value for x = 0.1 sample attains the highest value among all the samples for temperatures upto 30 K and fields upto 6T. For example the $J_c$ value at 5K for optimally doped sample (i.e. x = 0.1) is 8.4 x 10$^6$ A/cm$^2$ in self field, 1.1 x 10$^6$A/cm$^2$ at 3T and 5.3 x 10$^4$ A/cm$^2$ at 5T. On the other hand the $J_c$ value of pure sample is 2.4 x 10$^5$

A/cm$^2$ in self field, 1.3 x 10$^4$A/cm$^2$ at 3T and 2.9 x 10$^2$ A/cm$^2$ at 5T at 5K. Based on MgB$_2$ superconductor, most of devices can operate at 20K where as conventional superconductor can not operate due to low $T_c$.

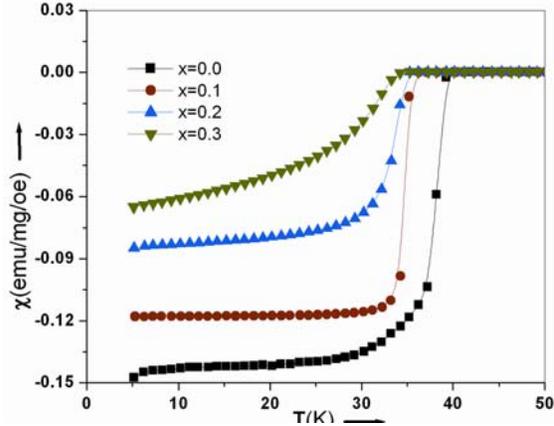

FIG.3. (Color) Temperature dependent dc magnetic susceptibility ($\chi$) behavior of MgB$_{2-x}$C$_x$ for x = 0.0, 0.1, 0.2 and 0.3.

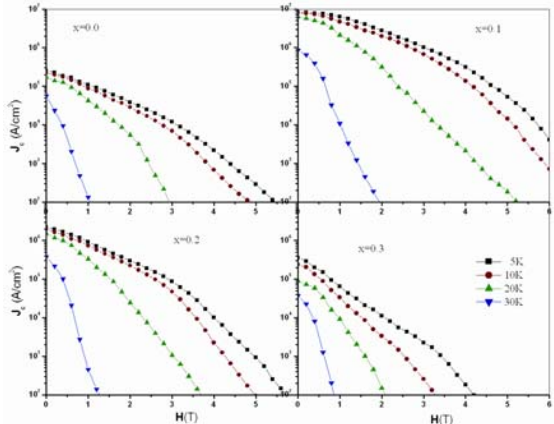

FIG.4. (Color) Critical current density $J_c$ as a function of applied magnetic field for MgB$_{2-x}$C$_x$ (a) x = 0.0, (b) x = 0.1, (c) x = 0.2 and (d) x = 0.3 at 5, 10, 20 and 30K.

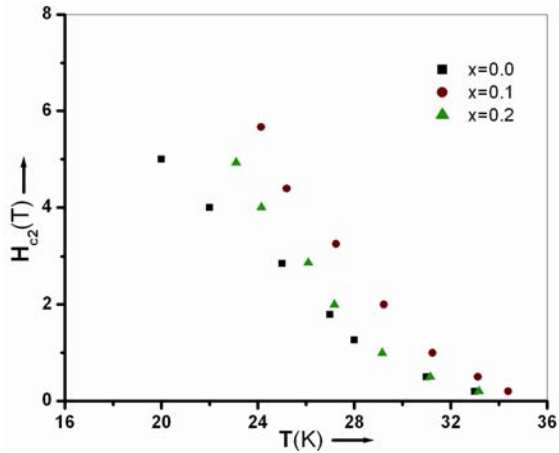

FIG.5. (Color) The upper critical field as a function of temperature for MgB$_{2-x}$C$_x$, x = 0.0, 0.1, 0.2 doped samples.

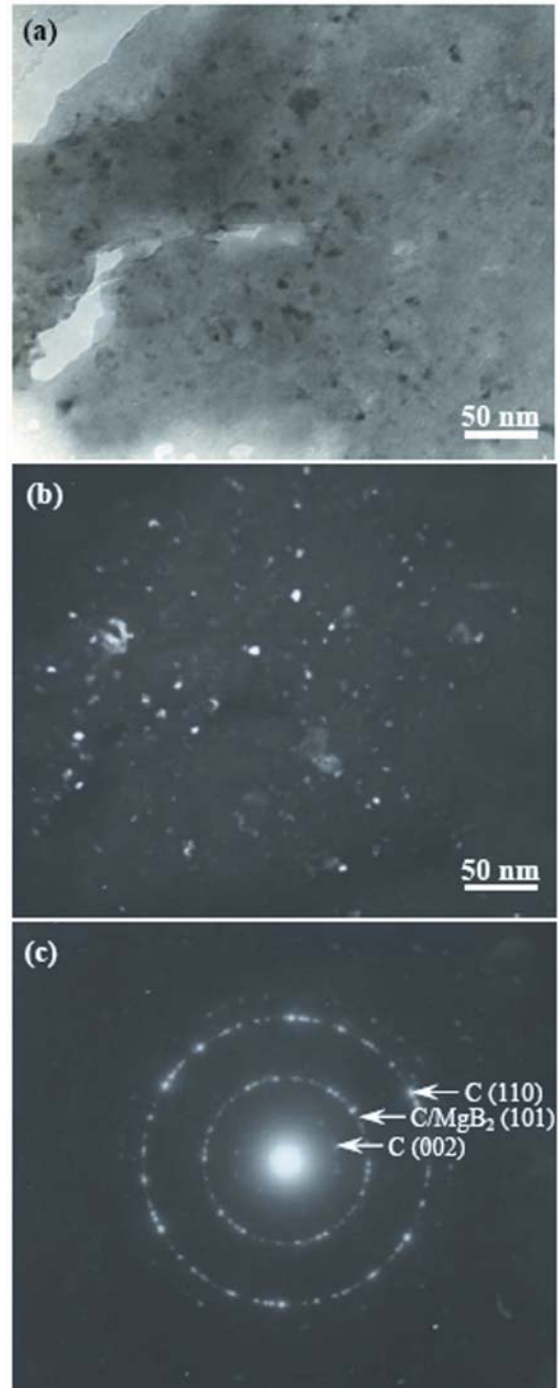

FIG.6. (Color) The representative TEM image of MgB$_{2-x}$C$_x$ with x = 0.1. (a) the bright field image shows presence of graphite nanoparticles (b) the dark field image of same region i.e. (a). This image also shows the presence of graphite nanoparticles in MgB$_2$ grain. (c) the selected area diffraction (SAD) pattern corresponding to image (a) shows spotty rings which have been identified as due to graphitic phase of carbon.

Therefore, $J_c$ value of optimally doped MgB$_2$ sample x = 0.1 at 20K is 3.2 x10$^5$, 2.3 x10$^4$ and 2.2 x10$^3$ A/cm$^2$ at 2T, 3T and 4T respectively. These values are significantly higher as compared to recent work done on MgB$_2$ [34,35,40-41]. The above values clearly show that doping has resulted in enhancement of $J_c$ for all fields.

We have done the isofield magnetization measurements as a function of temperature between 5 to 50K and noted the superconducting transition temperature at field 0.2, 0.5, 1, 2, 3, 4 and 5T. These superconducting transition temperature values are used to plot the $H_{c2}(T)$ versus temperature on horizontal axis as shown in Fig (5). The extrapolation of curve gives the $H_{c2}$ value at 0K. The value of $H_{c2}$ at 0K for pure MgB$_2$ sample is 16T and for x = 0.1 and x = 0.2 MgB$_2$ samples are 33T and 23.1T respectively. These values are also close to the values obtained by Werthamer Helfand-Hoheberg Model [42].

$$H_{c2}(0) = 0.7 T_c \left[ \frac{dHc2}{dT} \right]$$

which yields 17T, 31T and 25T as $H_{c2}(0)$ values for x = 0, x = 0.1 and x = 0.2 respectively. The enhancements of $H_{c2}$ suggest that the doping of graphite in MgB$_2$ induces disorder and results in shortening of electronic mean free path. The selective tuning of impurity scattering may improve the $H_{c2}$ value of MgB$_2$ superconductor. These values are comparable to recent work done by Pallecchi et al [43]. They have studied the effect of neutron irradiation on $H_{c2}$ values of MgB$_2$. The $H_{c2}$ values obtained in their experiment are also in agreement with the model of two band impurity [44,45]. It appears that graphite impurity in MgB$_2$ grain is enhancing the band scattering leading to increased $H_{c2}$ values.

Since the central aim of present investigation is to explore the enhanced superconducting properties of doped MgB$_2$ samples and their possible correlation with microstructural features, we have carried out investigations of microstructural features induced by doping of different concentration of doping in MgB$_2$. Fig. 6 (a) shows the bright field TEM micrograph for doped sample (x = 0.1). From this micrograph the presence of nanoparticles is easily discernible. Fig 6(b) shows dark field TEM image of same region [i.e. Fig. 6(a)] which confirms the presence of nanoparticles in MgB$_2$ grain. Fig. 6(c) shows the selected area diffraction (SAD) pattern corresponding to Fig. 6(a). The SAD pattern reveals the spotty rings which have been identified as due to graphitic phase of carbon. From Fig. 6(a, b & c) it can be easily concluded that the nanoparticles invariably present in MgB$_2$ grain are graphite nanoparticles. These particles are distributed homogeneously in MgB$_2$ grain. The size of nanoparticles has been found to be in the range of ~5-10 nm. It is interesting to note that some concentration of graphite have gone to honey comb layer on B site in MgB$_2$[30] (as it is evident from XRD analysis) and remaining concentration of graphite in the form of nanoparticles has been identified by TEM analysis. It may be pointed out that the microstructure of doped MgB$_2$ in the present case shows uniformly distributed graphitic nanoparticles only where as several other phases have been reported as inclusion particles by Yanwei et al [11]. This difference may be due to differences in method of sample preparation in two cases. Furthermore, the size and distribution of inclusion particles are more close to coherence length for MgB$_2$ in comparison to the inclusion of MgAg and LaB$_6$ reported in earlier studies [36,37]

In order to get broad insight of pinning mechanism, we have calculated values of $H_{irr}$ for undoped and doped MgB$_2$ samples using Kramer Scaling law [47]. We have plotted the values of $J_c^{0.5}H^{0.25}$ versus $H$ for temperatures 5, 10, 20 and 30 K. A straight line has been found at each temperature for $H$ ($H \geq 0.5$T) as can be seen in Fig. 7 (a,b&c). The values of $H_{irr}$ can be determined by extrapolating the straight line toward horizontal axis [2,5]. We have determined the value of $H_{irr}$ for different concentration of doping and undoped samples at different temperature as shown in fig. 7(d). We have used these values of $H_{irr}$ for further analyzing the shape of flux pinning force density $F_p$ versus reduced field $H^*$ ($H/H_{irr}$) curves.

Flux pinning mechanism associated with microstructural defects is assessed by analyzing the shape of $F_p$ curve as a function of applied field and temperature. It is expected that if pinning is arising due to grain boundary, $F_p$ will exhibit $H^{*0.5}(1- H^*)^2$ dependence [46]. For analyzing the shape of $F_p$ curve, the normalized pinning force $f = F_p/F_{pmax}$ is plotted against reduced field $H^*$ for different concentration of doping. The curve overlap, when a single pinning mechanism and centre is considered to be involved [47] i.e. grain boundaries alone act as pinning centres. Such scaling behaviour is commonly observed in Nb-based superconductors [48]. We have plotted f versus $H^*$ curves for pure and doped MgB$_2$ samples at temperature 5, 10, 20 and 30K as shown in Fig. (8). For the pure MgB$_2$ samples, the best fit is found for

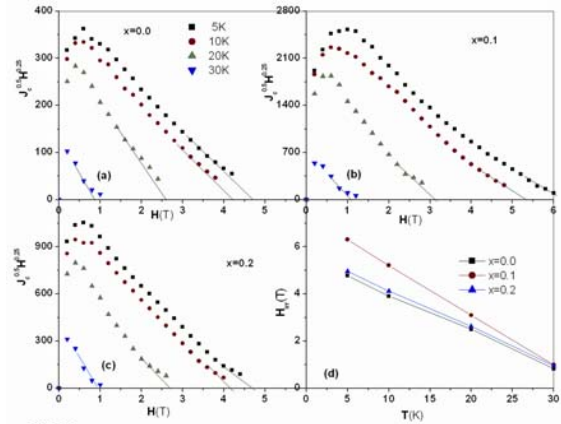

FIG.7. (Color) The variation of $J_c^{0.5}H^{0.25}$ with magnetic field H for (a) x = 0.0, (b) x = 0.1, (c) x = 0.2 at temperatures 5, 10, 20 and 30 K (d) shows the variation of $H_{irr}$ (it is deduced after extrapolating curves in (a), (b) & (c)) with temperature.

$H^{*0.5}(1- H^*)^2$ which is attributed to grain boundary pinning [46]. A similar observation has also been reported in pure polycrystalline bulk MgB$_2$ sample[2]. For x = 0.1 doped sample, the shape of the curve has significantly broadened and peak is shifted to higher $H^*$ value. This indicates that the nanoparticles inclusion has provided extra pinning centres [36]. Colley et al have found similar shift for carbon doping in $MgCNi_3$ superconductor and they have attributed this shift to core pinning by carbon nanoparticles [49]. When graphite concentration is high i.e. x > 0.1 the flux pinning force behaviour is suppressed. However, it may be pointed out that the result on the types of pinning based on figures 7 & 8 has some ambiguity. It only broadly indicates that the optimally doped $MgB_2$ sample shows high pinning strength at larger reduced magnetic fields.

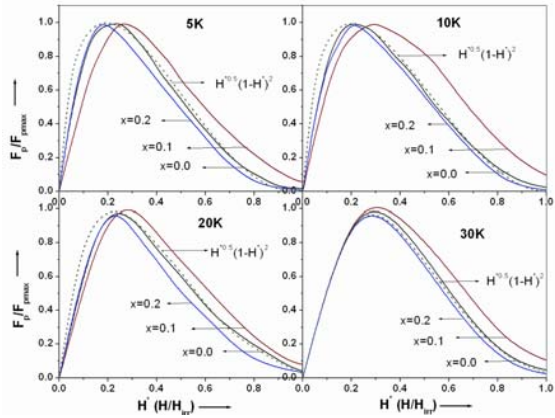

FIG.8. (Color) Normalized pinning force $F_p/F_{pmax}$ as a function of reduced magnetic field $H/H_{irr}$ at (a) 5K, (b) 10K, (c) 20K and (d) 30K of $MgB_{2-x}C_x$, for x = 0.0, 0.1, 0.2

## CONCLUSION

Based on the above results it can be concluded that we have synthesized successfully graphite doped $MgB_2$ superconductor by solid state reaction method employing encapsulation technique. The lattice parameters have changed noticeably due to the substitution of C atom in honeycomb layer of B in $MgB_2$ crystal. For optimally doped (x = 0.1) $MgB_2$ significant enhancement in the superconducting properties such as $J_c$, $H_{c2}$ and $H_{irr}$ have been found. Despite of some ambiguity in the shape of flux pinning force density curve, it may be concluded that the flux pinning strength is enhanced at higher magnetic fields in optimally graphite doped sample. TEM microstructures clearly show the presence of graphite nanoparticles (size ~ 5-10 nm) embedded in the $MgB_2$ grains. It may further be concluded that these nanoparticles having size comparable to the coherence length (~ 5-6 nm) of $MgB_2$ superconductor are responsible for the effective flux pinning and consequently enhancing the superconducting properties.

## ACKNOWLEDGEMENTS
The authors are grateful to Prof. A.R. Verma, Prof. C.N.R. Rao, Prof. S.K. Joshi and Prof. A.K. Roychaudhary for fruitful discussion and suggestions. Financial supports from UGC, DST-UNANST and CSIR are gratefully acknowledged. One of the authors (C. S.) is thankful to UGC New Delhi, Government of India for awarding senior project fellowship.